# A Live Maxwell's Demon

Howard J. M. Hanley[a],* and Debra J. Searles[b,c]

[a] Applied Mathematics, Research School of Physics and Engineering, Australian National University, Canberra, ACT 2600, Ausralia; [b] Centre for Theoretical and Computational Molecular Science, The Australian Institute for Bioengineering and Nanotechology, The University of Queensland, Queensland 4072, Australia; [c] School of Chemistry and Molecular Biosciences, The University of Queensland, Queensland 4072, Australia

**Abstract:**

A direct experimental replication of the Maxwell Demon thought experiment is outlined. The experiment determines the velocity/kinetic energy distribution of the particles in a sample by a novel interpretation of the results from a standard time-of-flight (TOF) small angle neutron scattering (SANS) procedure. Perspex at 293 K was subjected to neutrons at 82.2 K. The key result is a TOF velocity distribution curve that is a direct spatial and time-dependent microscopic probe of the velocity distribution of the Perspex nuclei at 293 K. Having this curve, one can follow the Demon's approach by selecting neutrons at known kinetic energies. One example is given: namely, two reservoirs - hot and cold reservoirs - were generated from the 293 K source without disturbing its original 293 K energy distribution.

**The Demon**

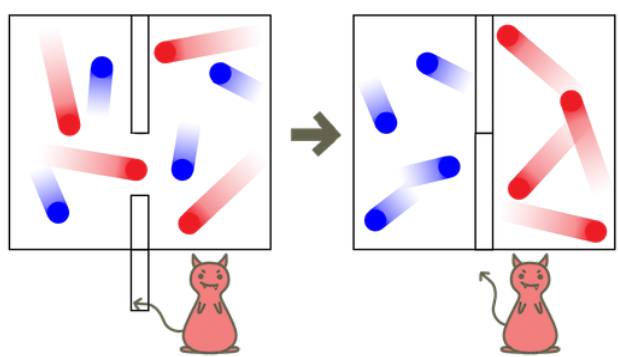

**Fig. 1.** Maxwell's Demon: *A hypothetical being imagined as controlling a hole in a partition dividing a gas-filled container into two parts, and allowing only fast-moving molecules to pass in one direction, and slow-moving molecules in the other [1] Image: A.T. Bernhardt.*

"Maxwell's Demon [1] lives on. After more than 130 years of uncertain life and at least two pronouncements of death, this fanciful character seems more vibrant than ever," so Harvey Leff and Andrew Rex remarked in 1990 [2]. Although the traditional debating point, that the Demon's results violate the second law of thermodynamics is no longer as controversial and provocative as it once was [3, 4], the Leff-Rex statement is still very true after another thirty years: a small sample of the relevant current literature is noted [5-12].

This note suggests a fresh and original approach in that we outline a real time, real space, direct attempt to follow the Demon's steps. We report on a time-of-flight (TOF) experiment using the first small angle neutron scattering spectrometer (AUSANS) to operate in Australia [13]. The procedure is standard [14-19] but the interpretation of the data is novel and is the core of the experiment.

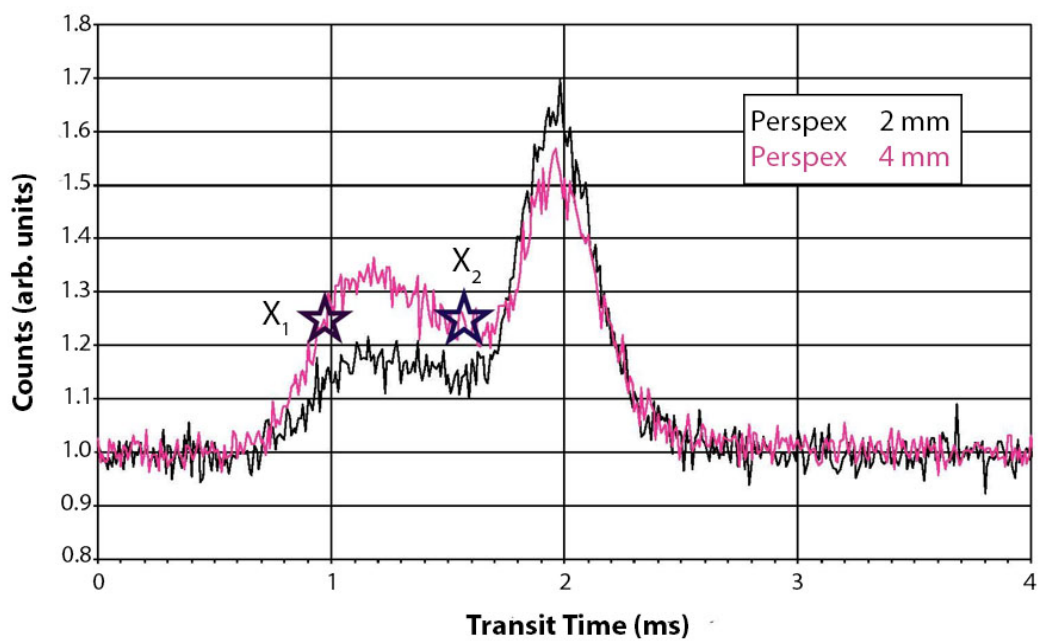

**Fig. 2.** *AUSANS time-of-flight spectrum of the scattering of incident neutrons at a kinetic temperature of 82.2 K passing through a Perspex sample at 293 K. The TOF of the incident beam is ~2 ms. The TOF of the secondary peak is ~ 1.2 ms which corresponds to a scattered neutron velocity, or kinetic temperature, close to that of the sample, 293 K [20,21].*

Incident neutrons at a kinetic temperature of 82.2 K were pulsed with a chopper, passed through a Perspex sample at room temperature, 293K and recorded on a $He^3$ 0.64 m x 0.64 m detector [13]. Figure 2 shows the distribution spectrum from the incident neutron beam, with a peak intensity at ~ 2 ms, and a secondary scattered spectrum with a peak intensity at ~ 1.2 ms [20, 21] In the specific context here, only this secondary spectrum is relevant because it is not the objective in this note to analyse the thermodynamic distribution in detail; that is a topic for future work. The key point is that the TOF data reflect pseudo-elastic neutron scattering from Perspex-nuclei collisions at 293 K [19]. The collisions of individual neutrons with these Perspex nuclei occur over extremely short times and, further most only collide with a single nucleus. Thus, the curve gives us a 'visual' microscopic probe of the velocity distribution of the neutrons that match the velocity distribution of the Perspex nuclei; compare with the Demon who looks into the velocity distribution of the gas molecules in his box (Fig. 1). Further, *a la* the Demon, we can identify those neutrons – strictly a distribution of neutrons (Fig. 2) – which are either more or less energetic with respect to an average. To give an arbitrary example: Fig. 2 shows two randomly chosen segments of the distribution indicated by the stars at positions $X_1$ and $X_2$. Merely, therefore, by selecting segments from the overall curve of the scattered neutrons we can locate a distribution of hot neutrons, and a distribution of cold neutrons, without altering the overall kinetic temperature distribution of the whole. In one way, we are ahead of the Demon who had to designate the location of the hotter with respect to the colder molecules by using a partition. But the end result is the same. The next step of the Demon's thought procedure could be to extract selected hot or cold molecules from the appropriate chamber, Fig. 1, and generate the corresponding hot and cold reservoirs. We can do this. The sets of the neutrons corresponding to those at positions $X_1$ and $X_2$ (Fig.2) can be binned separately because the neutrons will arrive at spatially different positions and times on the detector identified by their respective pathlengths. Furthermore, we should be able to detect a displacement of the scattering patterns between the two detector images simply because of the influence of gravity [22]. In short, we can spatially separate hot and cold neutrons to match the Perspex's nuclei's energy distribution and thus create two energy reservoirs from a single source. A simple calculation [20, 21] of the scattered neutrons at positions $X_1$ and $X_2$ of Fig. 2, indicates that the temperatures of the hot and cold reservoirs are ~ 329 K and ~ 130 K, respectively.

## Acknowledgements:
We are most grateful to have had Robert Knot as the collaborator in the early days of this study. We also acknowledge Stephen Williams, Vincent Craig, Adrian Sheppard, Andrew Kingston, Wilfred Fullagar, Liliana De Campo and Chris Garvey for their most constructive comments.

## References and Notes:

1. S. Ratcliffe (Ed), *Oxford Treasury of Sayings and Quotations* (Oxford University Press, 2011*)*; James Clerk Maxwell. *Theory of Heat* (Dover Publications Inc., New York 2001).
2. H. S. Leff, A. F. Rex (Eds), *Maxwell's Demon: Entropy, Information, Computing*

(Princeton Series in Physics, Princeton 1990); *Maxwell's Demon 2: Entropy, classical and Quantum Information, Computing* (Institute of Physics, Bristol 2003).

3. L. Szilard, Chap 3.1, p 124 in *On the Decrease of Entropy in a Thermodynamic System by the intervention of Intelligent Beings* (Princeton Series in Physics, Princeton. 1990).

4. J. Maddox, Maxwell's Demon: Slamming the Door. *Nature* **417**, 903 (27 June, 2002); C H. Bennett, Demons, Engines, and the Second Law. *Scientific American*. **257**, 108–117 (1987).

5. L. Zyga, Could Maxwell's Demon Exist in Nanoscale Systems? *Phys.org*. **24**, 1-9 (June, 2009).

6. P. Eshuis, K. van der Weele, D. Lohse, D. van der Meer, Experimental Realization of a Rotational Ratchet in a Granular Gas, *Phys. Rev. Lett.*, **104**, 248001- 4, (2010).

7. S. Toyabe, et. al. Experimental demonstration of information-to-energy conversion and validation of the generalized Jarzynski equality, *Nature Physics,* **6**, 988-992 (2010).

8. S. Vaikuntanathan, C. Jarzynski, Modeling Maxwell's demon with a microcanonical Szilard engine. *Phys. Rev, E.* **83**, 061120-9 (2011).

9. Z. Lu, D. Mandal, C. Jarzynski, Engineering Maxwell's demon*, Physics Today*, 60-61, (August 2014).

10. P.A. Camati, et. al, Experimental Rectification of Entropy Production by Maxwell's Demon in a Quantum System. *Phys. Rev. Lett.* **117**, 240502-5 (2016).

11. C. Nathanaël, et. al. Observing a quantum Maxwell demon at work. *PNAS.* **114** (29) 7561-7564 (2017).

12. M. Naghiloo, et. al. Information Gain and Loss for a Quantum Maxwell's Demon, *Phys. Rev. Lett*. **121**, 030604 – (2018**).**

13. R. B. Knott *Neutron News*, **9** *(1),* 23-32 (1998): H. J. M. Hanley, R. B. Knott, paper presented at the Australasian Soft Matter Scattering Workshop, RMIT, Melbourne, Feb. 11- 12 2016.

14. C. G. Windsor, *Pulsed Neutron Scattering* (Taylor and Francis, 1981).

15. J. R. D. Copley, T. J. Udovic, Neutron Time-of-Flight Spectroscopy. *J. Res. Natl. Inst. Stand. Technol*. **98** *(1*), 71–87 (1993).

16. C. Doa, et. al. Understanding inelastically scattered neutrons from water on a time-of-flight small-angle neutron scattering (SANS) instrument, *web site* arXiv:1309.1509 [physics.ins-det] (2013).

17. A. Sokolova, A. *Neutron News* **27**, 9-13 (2016).

18. A. R. Rennie, R. K. Heenan, Effects of Elastic Scattering in the Measurement and Analysis of SANS Data, International seminar on structural investigations on pulsed neutron sources. *Proceedings, Russian Federation: JINR,* 254-260 (1993).

19. R. E. Ghosh & A. R. Rennie, Assessment of Detector Calibrations Materials for SANS Experiments, *Appl. Crystallography*. **32**, 1157-1163 (1999).

20. Given the wavelength ($\lambda$) of a sample of mass (m); the energy (E), the velocity (v), and the temperature (T) are connected through the basic relations: $\lambda = h/(mv)$; $E = h^2/(2m\lambda^2)$; $T = E/k_B$ and $v^2 = 2E/m$. Here: m is the neutron mass ($1.674026 \times 10^{-27}$ kg); h, Planks constant; and $k_B$ Boltzmann's constant.

21. Taking the input beam wavelength (0.34 nm) as given, corrections to the data were determined by scaling the scattered neutron velocities with respect to the velocity ($0.34 = h/(mv)$): thus $v_{0.34} = 1165$ ms$^{-1}$. Further, assuming the scattered peak (Fig.2) is from the sample at 293 K, the equivalent scattered neutron velocity is 2200 ms$^{-1}$ leading to a scaled sample/input beam velocity ratio, $R_{velo} = 1.9$. The observed equivalent transit time/input beam ratio, $R_{time} \sim 2/1.2 \sim 1.7$: a ratio within 10% of $R_{velo}$  We thus determine the velocity of scattered neutrons at a selected transition time by assuming that $R_{time} = R_{velo}$. The uncertainty of the input beam wavelength is ~ 1.5%. Overall, we consider the calculated neutron velocities to have a conservatively estimated error of 12%. The source data for Fig. 2 are available on request.

22. See, for example: E. Aza, et. al.  Neutron beam monitoring for time-of-flight facilities with gaseous detectors, *Nuclear Instruments and Methods in Physics Research* A ***806***, 14–20 (2016).